\documentclass[pra,twocolumn,showpacs,superscriptaddress,amsfonts,amsmath,
floatfix]{revtex4}
\usepackage{graphicx}
\usepackage{psfrag}
\begin{document}
\title {Reply to arXiv:1002.4366, "Comment on 'Motion of an impurity in an ultracold quasi-one-dimensional gas of hard-core bosons'",
by S. Giraud and R. Combescot}
\author{M. D. Girardeau}
\email{girardeau@optics.arizona.edu}
\affiliation{College of Optical Sciences, University of Arizona, 
Tucson, AZ 85721, USA}
\author{A. Minguzzi}
\email{anna.minguzzi@grenoble.cnrs.fr}
\affiliation{Universit\'{e} Joseph Fourier, Laboratoire de Physique et 
Mod\'elisation des Mileux Condens\'es, C.N.R.S., B.P. 166, 38042 Grenoble, 
France}
\date{\today}
\begin{abstract}
In their Comment \cite{GirCom10} Giraud and Combescot point out that the contribution to the impurity-boson distribution function $\rho_{bi}(x-y)$
of a term we dropped is not negligible, rather than being negligible in the thermodynamic limit as we had conjectured. We now agree with them, but nevertheless our results for $\rho_{bi}$ are highly accurate for large impurity-boson mass ratio $m_i/m$ and remain qualitatively correct for all values of $m_i/m$ and all values of the boson-impurity coupling constant. 
\end{abstract}
\maketitle
This is our response to a recent Comment \cite{GirCom10} by S. Giraud and R. Combescot, 
which comments on our recent paper \cite{GirMin09} on
a moving impurity particle in an ultracold quasi-one-dimensional gas of hard-core bosons. The paragraph below is the verbatim
text of an Erratum 
resubmitted to PRA on February 22, 2010, 
initially submitted in July 2009 and placed on hold by PRA. 

In our derivations we dropped a momentum fluctuation term $\hat{p}_F^2$ in Eq. (6) for the Hamiltonian in the rest system of the
impurity, arguing, on the basis of a plausibility argument in Sec. V, that its contribution to the impurity-boson distribution
function should be negligible in the thermodynamic limit. As a consequence, one would conclude that the impurity-boson distribution function is independent of the impurity mass ratio $m_i/m$.
However, comparison with the solution of McGuire  \cite{McG65}  for the special case
of equal boson and impurity masses shows different behaviour.
For example, in the limit  where the coupling strength $\tilde \lambda= m g/k_F$ is equal to infinity, analytical expressions for the boson-impurity distribution function are available and  one respectively gets $\rho_{bi}(x)=1-|j_0(k_F x)|^2$ for the case of $m_i/m=1$  \cite{McG65}   and   $\rho_{bi}(x)=1-j_0(2 k_F x)$ for the case $m_i/m\to \infty$ \cite{GirWri00,GirMin09}.
This implies that the omitted momentum fluctuation term is not negligible; we thank S. Giraud and R. Combescot for pointing
this out. Therefore, our results for the impurity-boson distribution function are only approximate. The sentence
"This yields the exact ground state (total linear momentum $q=0$) and exact boson-impurity distribution function..." in the Abstract
should therefore be replaced by "In an approximation which neglects the fluctuation of linear momentum of the Bose gas, we find the
ground state  (total linear momentum $q=0$) and boson-impurity distribution function for arbitrary $m_i/m$ and arbitrary impurity-boson
interaction strength.". Also, the first sentence of Sec. V should be replaced by "Here we examine the contribution of the term
$\frac{\hat{p}_F^2}{2m_i}$ in Eq. (6).", and the next to last sentence of Sec. V should be omitted. The impurity-boson distribution function which we derived is  highly accurate for large impurity-boson mass ratio, i.e., $m_i/m\gg 1$, as previously shown in the
limit $m_i/m=\infty$ \cite{GirWri00} and recently confirmed by Giraud and Combescot \cite{GirCom09}. Comparison of the two analytical solutions mentioned above 
in the case where the difference is expected to be largest (very large impurity-boson coupling constant) indicates that  our solution 
remains qualitatively correct for all values of $m_i/m$ and all values
of the boson-impurity coupling constant, while not reproducing the details of the spatial behaviour of the impurity-boson distribution on the short-distance scale $k_Fx\le 1$. 
%
%
\acknowledgments
We are indebted to S. Giraud and R. Combescot for helpful correspondence which motivated our submission of an Erratum in July 2009. 
\end{document}